\documentclass[12pt,psfig]{article}

\usepackage{psfig}

\usepackage{scicite}

\usepackage{times}

\newenvironment{sciabstract}{%
\begin{quote} \bf}
{\end{quote}}

\newcounter{lastnote}

\title{A periodically-active pulsar giving insight into magnetospheric
physics}

\author
{M.~Kramer$^1$, A.~G.~Lyne$^1$, J.~T.~O'Brien$^1$, 
C.~A.~Jordan$^1$, \\ D.~R.~Lorimer$^1$ \\
\normalsize{$^{1}$Jodrell Bank Observatory, 
University of Manchester, Macclesfield, SK11~9DL, UK}
}

\date{Published in Science, Vol.~312, 549--551}

\begin{document} 

% Double-space the manuscript.

\baselineskip24pt

% Make the title.

\maketitle

% Place your abstract within the special {sciabstract} environment.

\begin{sciabstract}
PSR B1931+24 (J1933+2421) behaves as an ordinary isolated radio pulsar
during active phases that are 5-10 days long. However, the radio emission
switches off in less than 10 seconds and remains undetectable for the
next 25-35 days, then it switches on again. This pattern repeats 
quasi-periodically. The origin of this behaviour is unclear. Even more
remarkably, the pulsar rotation slows down 50\% faster when it is on
than when it is off. This indicates a massive increase in
magnetospheric currents when the pulsar switches on, proving that
pulsar wind plays a substantial role in pulsar spin-down. This allows
us, for the first time, to estimate the currents in a pulsar
magnetospheric during the occurrence of radio emission.
\end{sciabstract}

% In setting up this template for *Science* papers, we've used both
% the \section* command and the \paragraph* command for topical
% divisions.  Which you use will of course depend on the type of paper
% you're writing.  Review Articles tend to have displayed headings, for
% which \section* is more appropriate; Research Articles, when they have
% formal topical divisions at all, tend to signal them with bold text
% that runs into the paragraph, for which \paragraph* is the right
% choice.  Either way, use the asterisk (*) modifier, as shown, to
% suppress numbering.

%\section*{Introduction}

Pulsar radio emission is generally understood in terms of beams of coherent
plasma radiation from highly relativistic particles above the magnetic pole of
a rotating neutron star, producing pulses as the beam crosses Earth. However,
there is no satisfactory theory that explains the radio emission or even the
magnetospheric conditions that determine whether a neutron star emits at radio
wavelengths. Observationally, the typical active lifetime of a radio pulsar is
estimated to be about 10 million years during which, on long time scales,
pulsar emission is essentially steady ({\sl 1}). It was therefore surprising
when a unique activity pattern was found for the pulsar PSR~B1931+24 (also
known as PSR J1933+2421) during routine pulsar timing observations with the
76-m Lovell Telescope at the Jodrell Bank Observatory in the United Kingdom.

The pulsar had been considered to be a seemingly ordinary pulsar, with a spin
period of 813 ms ({\sl 2}) and a typical rotational frequency derivative of
$\dot\nu=-12.2\times 10^{-15}$ Hz~s$^{-1}$ (Table 1, {\sl 3}).  It was noted
that it exhibits considerable short-term rotational instabilities intrinsic to
the pulsar, known as timing noise, but shows no evidence indicating the
presence of any stellar companion.  It became clear a few years ago that the
pulsar was not detected in many of the regular observations and that the flux
density distribution was bimodal, the pulsar being either on or off.  Figure~1
shows the best sampled data span which covers a 20-month period between 1999
and 2001 and demonstrates the quasi-periodic pattern of the on-off sequences.
The power spectrum of the data reveals a strong $\sim35$-d periodicity with
two further harmonics, which reflect the duty cycle of the switching pattern
(Fig.~1). Studying a much longer time series from 1998 to 2005, including some
intervals of less densely sampled data, we find that the periodicities are
persistent but slowly vary with time over a period ranging from 30 to 40 days.
No other known pulsar behaves this way.

To investigate the nature of the variations, we examined the rotation rate of
the pulsar over a 160-day period (Fig.~2a). The variation is dominated by a
decrease in rotational frequency which is typical for pulsars.  However,
inspection of the longer sequences of the available data on the on phases in
the diagram reveals that the rate of decrease is even more rapid during these
phases, indicating values of the rotational frequency first derivative thatg
are higher than the average value.  This suggests a simple model in which the
frequency derivative has different values during the off and on phases.  Such
a model accurately describes the short-term timing variations that are seen
relative to a simple long-term slow-down model (Fig.~2b).  Over the 160-day
period shown, the pulsar was monitored almost daily, so that the switching
times were well defined, and a model could be fitted to the data with good
precision. The addition of a single extra parameter (that is, two values of
the frequency derivative rather than one) reduced the timing residuals by a
factor of 20 and provides an entirely satisfactory description of the data. A
similar fitting procedure was applied to other well-sampled sections of the
data and produced consistent model parameters (Table~1), giving values for the
rotational frequency derivatives of $\dot\nu_{\rm off}=
-10.8(2)\times10^{-15}$ Hz s$^{-1}$ and $\dot\nu_{\rm on}=
-16.3(4)\times10^{-15}$ Hz s$^{-1}$.  These values indicate that there is a
$\sim$50\% increase in spin-down rate of the neutron star when the pulsar is
on.

The observed quasi-periodicity in pulsar activity and its time scale
have never been seen before as a pulsar emission phenomenon and are
accompanied by massive changes in the rotational slow-down rate. This
raises a number of questions.  Why does the emission switch on and
off?  Why is the activity quasi-periodic?  Why is the pulsar spinning
down faster when it is on?

On the shortest, pulse-to-pulse time scales, intrinsic flux density variations
are often observed in pulsar radio emission. The most extreme case is
displayed by a small group of pulsars that are known to exhibit ``nulls'' in
their emission; that is, the random onset of a sudden obvious lack of pulsar
emission, typically lasting between one and a few dozen pulsar rotation
periods ({\sl 4}).  An acceptable explanation for such nulling, which appears
to be the complete failure of the radiation mechanism, is still missing.  This
nulling previously represented the longest known time scales for an intrinsic
disappearance of pulsar emission. The facts that the off periods of PSR
B1931+24 are five orders of magnitude longer than typical nulling periods,
that the activity pattern is quasi-periodic and that not a single null has
been observed during on periods strongly suggest that the phenomenon found
here is different from nulling.

The approximate 35-day period might be attributed to free
precession, although we find no evidence of expected ({\sl 5})
profile changes.  Although switches between states are rare
events, we have been able to observe one switch from on to off that
occurred within 10 seconds, the time resolution being limited by the
signal-to-noise ratio of the observations.  The sudden change and the
quasi-periodicity point toward a relaxation oscillation of unknown
nature within the pulsar system, rather than precession.

What can cause the radio emission to cut off so quickly?  The energy
associated with the radio emission from pulsars accounts for only a
very small fraction of the pulsar's slow-down energy which may suggest
that the disappearance of radiation is simply due to the failure of
the coherence condition in the emission process ({\sl 6}). However,
in that case, the long timescales of millions of pulsar rotations are
hard to understand. One alternative explanation is that there is a
more global failure of charged particle currents in the magnetosphere.

The large changes in slow-down rate that accompany the changes in radio
emission can also be explained by the presence or absence of a plasma whose
current flow provides an additional braking torque on the neutron star. In
this model, the open field lines above the magnetic pole become depleted of
charged radiating particles during the off phases when the rotational slow
down $\dot{\nu}_{\rm off}$, is caused by a torque dominated by magnetic dipole
radiation ({\sl 7,8}). When the pulsar is on, the decrease in rotational
frequency, $\dot{\nu}_{\rm on}$, is enhanced by an additional torque provided
by the outflowing plasma, $T \sim \frac{2}{3c} I_{pc} B_0 R_{pc}^2$, where
$B_0$ is the dipole magnetic field at the neutron star surface and $I_{pc}\sim
\pi R_{pc}^2 \rho c$ is the electric current along the field lines crossing
the polar cap, having radius of $R_{pc}$ (e.g.~{\sl 9}). (In order to be
consistent with existing literature such as {\sl 9}, we quote formulae in
centimetre-gram-second-units but refer to numerical values in SI units.)  The
charge density of the current can be estimated from the difference in loss in
rotational energy during the on and off phases. When the pulsar is on, the
observed energy loss, $\dot{E}_{\rm on}=4\pi^2 I \nu \dot{\nu}_{\rm on}$, is
the result of the sum of the magnetic dipole braking as seen during the off
phases, $\dot{E}_{\rm off}=4\pi^2 I \nu \dot{\nu}_{\rm off}$, and the energy
loss caused by the outflowing current, $\dot{E}_{\rm wind}= 2\pi T\nu$;
that is,~$\dot{E}_{\rm on}=\dot{E}_{\rm off}+\dot{E}_{\rm wind}$ where $I$ is the
moment of inertia of the neutron star.  From the difference in spin-down rates
between off and on phases, $\Delta \dot{\nu}=\dot{\nu}_{\rm
  off}-\dot{\nu}_{\rm on}$, we can therefore calculate the charge density
$\rho=3I\Delta \dot{\nu}/R_{pc}^4 B_0$ by computing the magnetic field $B_0=
3.2\times10^{15}\sqrt{-\dot{\nu}_{\rm off}/\nu^3}$ Tesla and the polar cap
radius $R_P=\sqrt{2\pi R^3\nu/c}$ for a neutron star with radius $R=10$km and
a moment of inertia of $I=10^{38}$ kg m$^2$ ({\sl 10}). We find that the
plasma current that is associated with radio emission carries a charge density
of $\rho=0.034$ C m$^{-3}$. This is remarkably close to the charge density
~$\rho_{\rm GJ} =B_0\nu/c$ in the Goldreich-Julian model of a pulsar
magnetosphere ({\sl 11}); that is~$\rho_{\rm GJ} =0.033$ C m$^{-3}$.

Such charge density is sufficient to explain the change in the neutron
star torque, but it is not clear what determines the long timescales
or what could be responsible for changing the plasma flow in the
magnetosphere. In that respect, understanding the cessation of
radiation that we see in PSR~B1931+24 may ultimately also help us to 
understand ordinary nulling.  Whatever the cause is, it is conceivable
that the onset of pulsar emission may be a violent event that may be
revealed by high-energy observations. Although an archival search for
x-ray or $\gamma$-ray counterparts for PSR~B1931+24 has not been
successful, the relatively large distance from the pulsar ($\sim 4.6$
kpc) and arbitrary viewing periods may make such a detection unlikely.

The relation between the presence of pulsar emission via radiating
particles and the increased spin-down rate of the neutron star
provides strong evidence that a pulsar wind plays a substantial role
in the pulsar braking mechanism. Although this has been suggested in the
past ({\sl 12}), direct observational evidence has been
missing so far. As a consequence of the wind contribution
to the pulsar spin-down, magnetic fields estimated for normal pulsars
from their observed spin-down rates are likely to be overestimated.

The discovery of PSR~B1931+24's behaviour suggests that many more such
objects exist in the Galaxy but have been overlooked so far because
they were not active during either the search or confirmation
observations. The periodic transient source serendipitously found
recently in the direction of the galactic centre ({\sl 13}) may turn
out to be a short-timescale version of PSR~B1931+24 and hence to be a
radio pulsar.  In general, the timescales involved in the observed
activity patterns of these sources pose challenges for observations
scheduled with current telescopes. Instead, future telescopes with
multibeaming capabilities, like the Square-Kilometre-Array or the Low
Frequency Array, which will provide continuous monitoring of such
sources, are needed to probe such timescales which are still almost
completely unexplored in most areas of astronomy.

%\paragraph*{In-line math.}  The utility that we use for converting

% Your references go at the end of the main text, and before the
% figures.  For this document we've used BibTeX, the .bib file
% scibib.bib, and the .bst file Science.bst.  The package scicite.sty
% was included to format the reference numbers according to *Science*
% style.

%\bibliographystyle{Science}
%\bibliography{journals,modrefs,psrrefs,crossrefs}

% Following is a new environment, {scilastnote}, that's defined in the
% preamble and that allows authors to add a reference at the end of the
% list that's not signaled in the text; such references are used in
% *Science* for acknowledgments of funding, help, etc.

%\begin{scilastnote}
%\item
%\end{scilastnote}

% For your review copy (i.e., the file you initially send in for
% evaluation), you can use the {figure} environment and the
% \includegraphics command to stream your figures into the text, placing
% all figures at the end.  For the final, revised manuscript for
% acceptance and production, however, PostScript or other graphics
% should not be streamed into your compliled file.  Instead, set
% captions as simple paragraphs (with a \noindent tag), setting them
% off from the rest of the text with a \clearpage as shown  below, and
% submit figures as separate files according to the Art Department's
% instructions.

\clearpage

\noindent {\bf Table 1} Observed and derived parameters of PSR~B1931+24.
Standard
($1\,\sigma$) errors are given in parentheses after the values and are
in units of the least significant digit. The distance is estimated
from the dispersion measure and a model for the interstellar free
electron distribution ({\sl 14}).
Definitions for characteristic age, surface magnetic field and
spin-down luminosity can be found in~{\sl lk05}.

\begin{table*}[ht]
%\begin{scriptsize}
\begin{center}
\begin{tabular}{lc}
\hline
 & \\ 
Right ascension (J2000) & $19^{\rm{h}}33^{\rm{m}}37^{\rm{s}}.832(14)$ \\
Declination (J2000) & $+24^\circ 36' 39''.6(4)$ \\
Epoch of frequency (MJD) & 50629.0 \\
Rotational frequency $\nu$ (Hz) & 1.2289688061(1) \\
Rotational frequency derivative $\dot\nu$ (Hz s$^{-1}$) & $-12.2488(10) \times 10^{-15}$ \\
Rotational frequency derivative on $\dot{\nu}_{\rm on}$ (Hz s$^{-1}$) & $-16.3(4) \times 10^{-15}$ \\
Rotational frequency derivative off $\dot{\nu}_{\rm off}$ (Hz$^{-1}$) & $-10.8(2) \times 10^{-15}$ \\
Dispersion measure DM (cm$^{-3}$pc) & 106.03(6) \\
Flux density during on phases at 1390 MHz ($\mu$Jy) & 1000(300) \\
Flux density during off phases at 1390 MHz ($\mu$Jy) & $\leq$2 \\
Flux density during on phases at 430 MHz ($\mu$Jy) & 7500(1500) \\
Flux density during off phases at 430 MHz ($\mu$Jy) & $\leq$40 \\
Active duty cycle (\%) & 19(5) \\
 & \\
Characteristic age $\tau$ (Myr) & 1.6 \\
Surface magnetic field strength $B$ (Tesla) & $2.6\times 10^{8}$ \\
Spin-down luminosity $\dot E$ (W) & $5.9\times10^{25}$ \\
Distance (kpc) & $\sim$4.6 \\
 & \\
\hline
\end{tabular}
\end{center}
%\end{scriptsize}
\end{table*}

\clearpage

\noindent {\bf Fig. 1.} 
Time variation of the radio emission of PSR~B1931+24.
During the on phases, the pulsar is easy to detect and has the
stable long-term intrinsic flux density associated with most normal
pulsars. Since 1998, the pulsar has been observed as frequently as
twice a day. a) A typical sequence of observations covering a 20-month
interval is indicated by the black lines. It
shows respectively the times of observation and the times when
PSR~B1931+24 was on. It is clear that the pulsar is  not
visible for $\sim$80\% of the time.  b) The appearance
of the pulsar is quasi-periodic nature, demonstrated by the
power spectrum of the intensity obtained from the Fourier Transform of
the autocorrelation function of the mean pulse flux density obtained
over the same 20-month interval. c) Histograms of the durations
of the on (solid) and off (hatched) phases.  In off phases,
integration over several weeks shows that any pulsed
signal has a mean flux density of less than than 2~$\mu$Jy at
1400~MHz. A deep observation with the Arecibo telescope on Puerto Rico
provides an upper flux density limit of $40\mu$Jy at 430
MHz. Simultaneous observations at frequencies between 430 and
1400 MHz show that the presence or absence of radio emission is a
broad-band phenomenon.

\clearpage

\noindent {\bf Fig. 2.} Variation of the rotational frequency of
PSR~B1931+24.  a) Evolution of the rotational frequency over a 160-day
period.  The errors in the measurement of the data points are smaller
than the size of the symbols.  The variation is dominated by the
long-term spin-down: a gradual decrease in rotational 
frequency. The best-fit straight-line through the points
is shown, representing a frequency derivative of
$\dot\nu=-12.2\times10^{-15}$ Hz s$^{-1}$.  However, an inspection of
the data reveals that,when the pulsar is on, the slope and hence the
magnitude of the derivative are even greater.  This suggests a model in
which the frequency derivative has different values during the off and
on periods.  b) Examining this possibility more closely, 
timing residuals, which are the differences between the observed pulse
arrival times and those derived from a simple long-term slow-down
model (Table~1), show substantial short-term variations.  Over
the period covered by this figure, the pulsar was monitored almost
daily, so that the off and on periods were well defined and the model
described above could be fitted to the data with good precision. The
fitted model is shown as a continuous line overlying the data points
and clearly describes the data very well.

\newpage
\centerline{\psfig{file=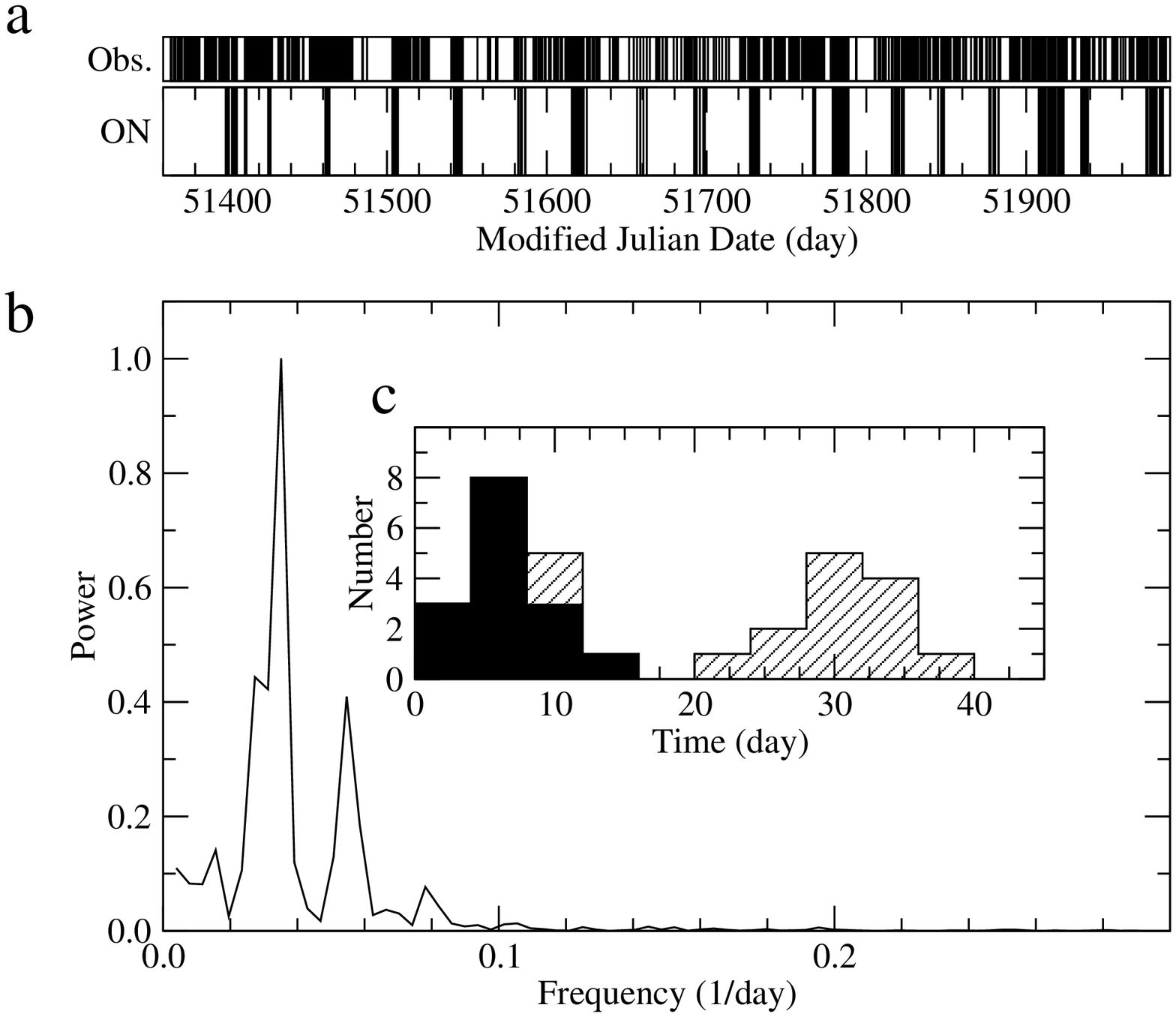,angle=0,width=14cm}\\ Fig.~1}

\newpage
\centerline{\psfig{file=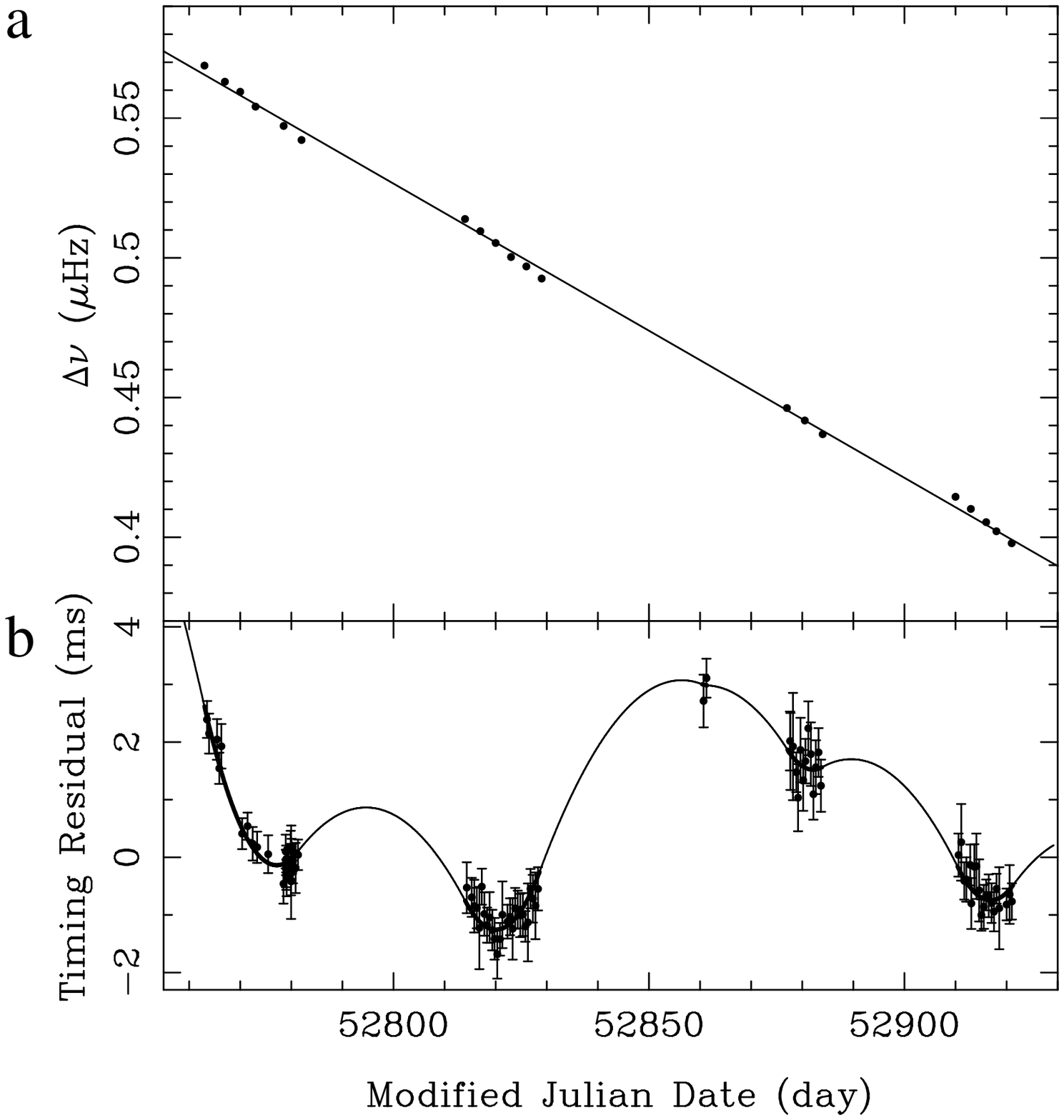,angle=0,width=14cm}\\ Fig.~2}

\end{document}